\documentclass[english,prl,twocolumn,floatfix,showpacs]{revtex4}
\usepackage{graphicx}  
\usepackage{dcolumn}   
\usepackage{bm}        
\usepackage{amssymb}   
\usepackage{amsmath}
\usepackage{color}

\usepackage[T1]{fontenc}
\usepackage[latin9]{inputenc}
\usepackage{amsthm}
\usepackage{color}
\usepackage{amsmath}
\usepackage{graphicx}

\makeatletter
\@ifundefined{textcolor}{}
{%
 \definecolor{BLACK}{gray}{0}
 \definecolor{WHITE}{gray}{1}
 \definecolor{RED}{rgb}{1,0,0}
 \definecolor{GREEN}{rgb}{0,1,0}
 \definecolor{BLUE}{rgb}{0,0,1}
 \definecolor{CYAN}{cmyk}{1,0,0,0}
 \definecolor{MAGENTA}{cmyk}{0,1,0,0}
 \definecolor{YELLOW}{cmyk}{0,0,1,0}
 }

\makeatletter

\newcommand{\Rmnum}[1]{\expandafter\@slowromancap\romannumeral #1@}
\makeatother

\newcommand{\ed}[1]{\textcolor{black}{#1}}
\newcommand{\edl}[1]{\textcolor{black}{#1}}

\newcommand{\be}{\begin{equation}}
\newcommand{\ee}{\end{equation}}

\newcommand{\rd}{{\rm d}}

\newcommand{\di}{{\sf DI}}

\def\lsim{\mathrel{\rlap{\lower4pt\hbox{$\sim$}}
    \raise1pt\hbox{$<$}}}                
\def\gsim{\mathrel{\rlap{\lower4pt\hbox{$\sim$}}
    \raise1pt\hbox{$>$}}}

\renewcommand\[{\begin{equation}}
\renewcommand\]{\end{equation}}
\usepackage[english]{babel}
\addto\captionsenglish{}
\makeatother

\usepackage{babel}

\begin{document}

\title{Persistence-driven durotaxis: Generic, directed motility in rigidity gradients}

\author{Elizaveta A. Novikova$^{1,4}$, Matthew Raab$^{2}$, Dennis E. Discher$^{3}$ and Cornelis Storm$^{4,5}$}

\affiliation{
$^{1}$ Institute for Integrative Biology of the Cell(I2BC), Institut de Biologie et de Technologies de Saclay(iBiTec-S), CEA, CNRS, Universite Paris Sud, F-91191 Gif-sur-Yvette cedex, France,
 $^{2}$ CNRS UMR144, Institut Curie, 12 rue Lhomond, 75005 Paris, France $^{3}$Molecular \& Cell Biophysics and Graduate Group in Physics, University of Pennsylvania, Philadelphia, Pennsylvania 19104, USA, $^{4}$Department of Applied Physics, $^{5}$Institute for Complex
Molecular Systems, Eindhoven University of Technology, P. O. Box 513,
NL-5600 MB Eindhoven, The Netherlands}

\date{\today}
\begin{abstract}
Cells move differently on substrates with different \ed{rigidities}: The persistence time of their motion is higher on stiffer substrates. We show that this behavior---in and of itself---results in a net flux of cells directed up a soft-to-stiff gradient. Using simple random walk models with varying persistence and stochastic simulations, we characterize the propensity to move in terms of the durotactic index also measured in experiments. A one-dimensional model captures the essential features and highlights the competition between diffusive spreading and linear, wavelike propagation. Persistence-driven durokinesis is generic and may be of use in the design of instructive environments for cells and other motile, mechanosensitive objects.
\end{abstract}


\maketitle

Cells are acutely aware of the mechanical properties of their surroundings. The rigidity, or lack thereof, of the substrate to which a cell is adhering determines a number of crucial processes: Differentiation, gene expression, proliferation, and other cellular \edl{decisions have} all been shown to be affected by the stiffness of the surrounding matrix \ed{\cite{EvansMinelli2009,yeung2005effects,Discher18112005,Engler2006677,Huck2012,justin2011stiffness,Pham2016}}. Cells also {\em move} differently depending on the \ed{rigidity} of the substrate. One of the more striking manifestations of this is the near-universal tendency of motile cells to travel up rigidity gradients in a process generally referred to as {\em durotaxis} \cite{lo2000cell,charras2014physical,pelham1997cell,vincent2013mesenchymal,Raab2012crawling,
missirlis2013combined,isenberg2009vascular,Maiuri2015374}, a term that emphasizes the similarity to {\em chemotaxis}, the ability of cells to move directedly in chemical gradients. Chemotaxis---generally believed to offer significant evolutionary advantage---allows cells, for instance, to move towards sources of nutrients. For durotaxis, such advantage is less obvious. Motion in stiffness gradients could allow neutrophils and cancer cells to seek out optimal locations for extravasation \cite{brabek2010role,jannat2010neutrophil,oakes2009neutrophil}, or stem cells to contribute to mitigation or regeneration of stiff scars and injured tissues \cite{dingal2015fractal}. Durotactic motion is universal: without exception it is {\em away from softer, towards stiffer}.
\begin{figure}[t!]
\includegraphics[width=.9\columnwidth]{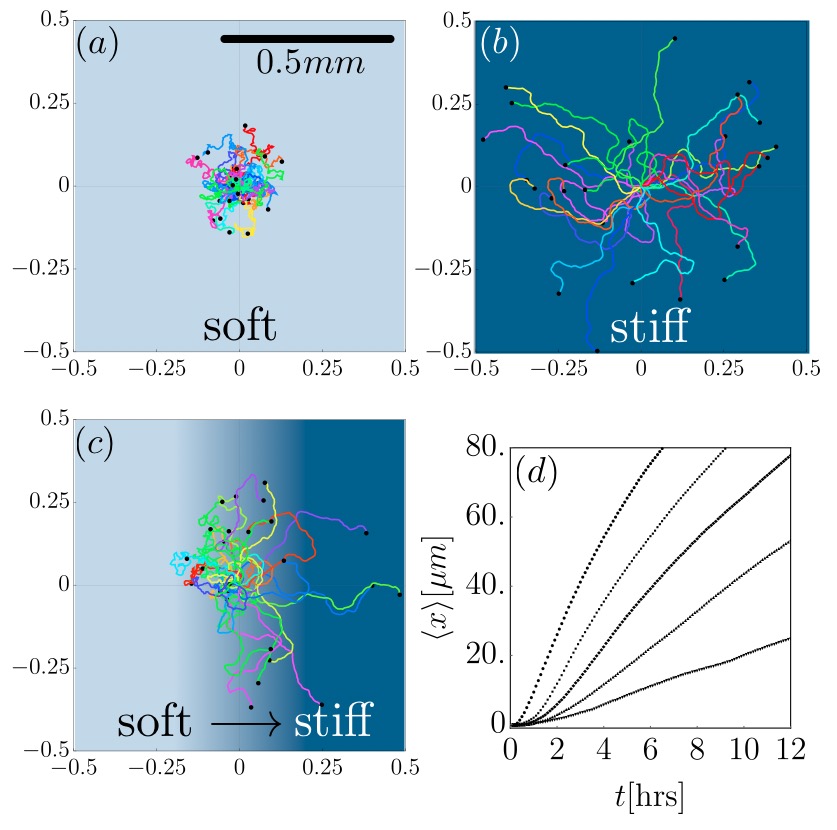}\caption{{\em Persistence-dependent motility}. Simulated trajectories \ed{(2D model)} of 25 cells, departing from the origin at $t=0$ with a linear velocity of 50 $\mu$m/hr. Total time is 12 hrs, cellular positions are recorded at 6-minute intervals. A black dot marks the end of each cell trajectory. (a) Cells on a soft substrate, with a low persistence time $\tau_p=0.2$ hrs. (b) stiff substrate; persistence time $\tau_p=$2 hrs. (c) Gradient substrate, with persistence time increasing linearly from 0.2 to 2 \ed{hrs} over the $x$-range $[-0.1,0.1]$ mm (i.e., $\Delta \tau_p/\Delta x= $ 9 hrs/mm). (d) averaged $x$-displacement \ed{in} the gradient region, for different \ed{widths of the gradient regions, and hence} the gradient steepnesses (top to bottom: $\Delta \tau_p/\Delta x= $ 90 hrs/mm, 18 hrs/mm, 9 hrs/mm, 4.5 hrs/mm, 1.8 hrs/mm).} \label{Fig1}
\end{figure}
In addition to an overall motion in a gradient region, the nature of cellular motion {\em itself} was shown to change quantitatively depending directly on the local rigidity of the substrate, with cells moving more persistently on more rigid substrates. \edl{In this Letter, we examine how locally different, persistent motility affects the global transport of cells}. We find that soft-to-stiff durotaxis is a necessary consequence of stiffness-dependent persistence, with or without a rigidity-dependent crawling speed. The mechanism we uncover  is fundamentally different from those reported in earlier theoretical works on durotaxis \cite{harland2011adhesion, stefanoni2011numerical}: the cells take no directional cues from the gradient region, but their persistence---a nondirectional property---is stiffness dependent. This experimentally established fact, alone, suffices to generate durotactic motion.

\begin{figure}[t!]
\includegraphics[width=\columnwidth]{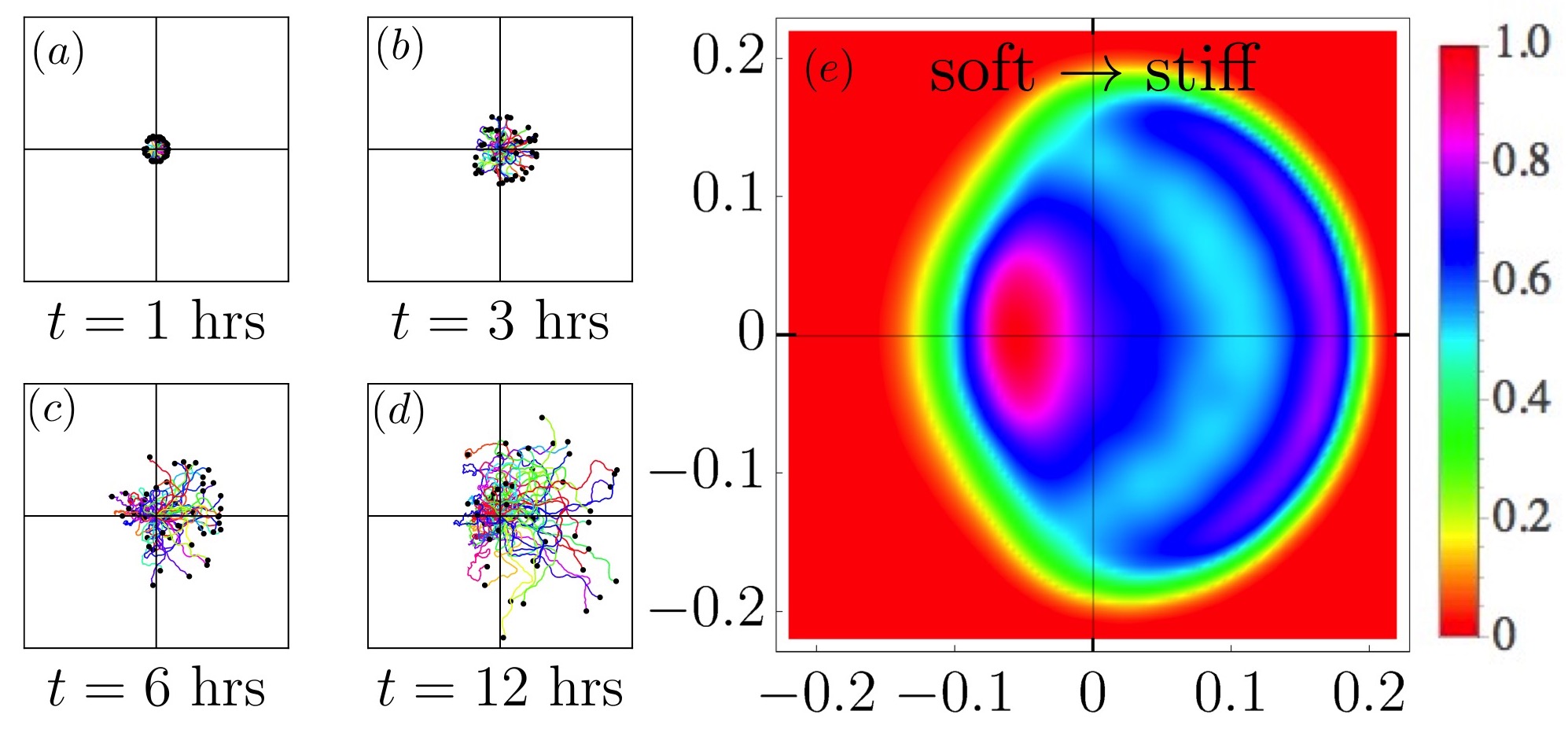}\caption{{\em Evolution of probability with time}. Simulated trajectories \ed{(2D model)} of 50 cells, departing from the origin at $t=0$ with a linear velocity of 50 $\mu$m/hr on a persistence gradient, increasing linearly from 0.2 to 2 \ed{hrs} over the $x$-range $[-0.1,0.1]$ mm (i.e., $\Delta \tau_p/\Delta x= $ 9 hrs/mm). The cells were tracked for 12 hrs, their positions recorded at 6-minute intervals. A black dot marks the end of each cell trajectory. (a)-(d) As time progresses, the asymmetry becomes increasingly clear. (e) The probability distribution ${\sf P}(x,y)$ (rescaled such that its maximal value is 1) at $t=4$ hrs clearly shows a double-peaked structure: a diffusive peak on the soft side, and a wavefront further out on the rigid side.} \label{Fig2}
\end{figure}

{\em Definitions and experimental observations.} For cells moving on uniformly \ed{rigid} substrates, most experiments record the paths of motile cells by tabulating, at fixed time intervals $\Delta t=t_{i+1}-t_i$, their position $\vec r(t_i)=\{x(t_i),y(t_i)\}$. The resulting time series constitutes a discrete-time Random Walk (RW). These cellular RW paths display a certain amount of {\em persistence}; the tendency to keep moving along the same direction (or, equivalently, the cell's inability to turn on very short timescales). This persistence is quantified by the persistence time $\tau_p$. For cells moving at a constant linear velocity $v_c$, this persistence time may be obtained by analyzing the displacement statistics of the path, either as the decay time \edl{of} the tangent autocorrelation, or by fitting to the mean squared displacement for a persistent random walk (PRW) \cite{rubinstein2003polymers}
\be\label{eq1}
\langle |\vec r^2| \rangle(t) = 2 v_c^2 \tau_p^2\left(\frac{t}{\tau_p}+e^{-t/\tau_p}-1\right)\,.
\ee
We note, that while the PRW correctly describes cellular motility in 2D, it fails in 3D \cite{Wu18032014}---one of many important differences between 2D and 3D processes of cellular adhesion and migration. We restrict ourselves to the case of 2D motility here, to make our general point. The limiting behavior of Eq.\ref{eq1} is instructive: for short times $t \ll \tau_p$ it describes ballistic motion $\langle |\vec r^2| \rangle(t)\approx (v_c t)^2$, whereas for long times $t \gg \tau_p$ the motion is a pure random walk; $\langle |\vec r^2| \rangle(t)\approx 2 v_c^2 \tau_p t$. Thus, the persistence time is the characteristic timescale for the crossover between ballistic and diffusive motion. A trivial point, which nonetheless bears repeating here, is that the first moment of the vectorial displacement vanishes, for RW and PRW alike: $\langle \vec r\rangle(t)=\vec 0$---this is no longer the case for durotactic processes. A meaningful question, now, is to ask how the parameters that quantify persistence and directed displacement {\em change} with the properties of the substrate. While the tendency to move from soft to stiff substrates has been broadly noted and characterized \cite{keller1971model,horstmann20031970,codling2008random,othmer1988models,mccutcheon1946vol}, the persistence of cells as they do so has only recently begun to be quantitatively addressed. A potential relation between the two has been hinted at in passing \edl{\cite{Raab2012crawling}}, but not further substantiated. In experiments recording the motility of fibroblasts on uniformly \ed{rigid} PEG hydrogels, Missirlis and Spatz \cite{missirlis2013combined} demonstrate that the persistence time, quantified by a Directionality Index \edl{$\Delta(t)=\sqrt{\langle |\vec r^2| \rangle}(t)/(v_c t)\propto (\tau_p/t)^{1/2}$} recorded at the same time on substrates coated with different ligands, rises by about a factor of 3 when the substrate stiffness is increased from 5.5 to 65.7 kPa. Over the same range of stiffnesses, a {\em decrease} of $v_c$ by about 33\% (from 60 $\mu$m/hr to 40 $\mu$m/hr) is reported. House {\em et al} \cite{house2009tracking} place fibroblasts on uniformly \ed{rigid} PAM hydrogels, and report that their persistence time increases by a factor of 3 when the gel stiffness is varied from 10 kPa to 150 kPa. Interestingly, and in contrast to Missirlis and Spatz, House {\em et al.} report an {\em increase} of $v_c$ with substrate stiffness by a factor of about 2 from 21.6 $\mu$m/hr to 42.7 $\mu$m/hr over the same stiffness range. A preliminary test, reported in \cite{house2009tracking}, suggests the cells move in the direction of increased persistence. In earlier work \cite{Raab2012crawling}, Raab {\em et al.} quantify the motility of mesenchymal stem cells on uniform PAM substrates - likewise showing an increase in persistence time of about a factor of 3 from 0.7 hrs to 2.1 hrs when the substrate stiffness is varied from 1 kPa to 34 kPa. Raab {\em et al.} report no significant change in the cell velocity $v_c$ over the entire range of stiffnesses they study. Importantly, however, Raab {\em et al.} also show that the same cells, on the same substrates that are now gradiented in stiffness from 1 kPa to 34 kPa, move towards the stiff side with a durotaxis index \edl{that over the course of about 2 hrs rises from 0 to 0.2}. In summary, experiments unanimously suggest that cells move more persistently on stiffer substrates, and that they move from soft to rigid. This behavior is independent of the relation between velocity and stiffness, which appears to be more cell-type dependent although a recent work suggests that speed and persistence may be correlated \cite{Maiuri2015374}. The empirical fact that two behaviors---increasing persistence and soft-to-stiff motion---coincide suggests they might not be independent. We now examine whether there is indeed a causation underlying the correlation.

{\em Simulation setup and results}. We consider a 2D substrate, endowed with a gradient in stiffness that manifests itself as a position-dependent persistence time $\tau_p(x)$ and a position-dependent velocity $v_c(x)$. To simulate the variable-persistence, variable cell speed PRW in the gradient region, we generate trajectories as follows: Starting in the origin at $t=0$, a random initial direction $\theta_0$ is chosen, along which the cell is displaced by a distance $\Delta r_1=v_c(0) \Delta t$. For all subsequent steps, a deviation angle $-\pi<\delta\theta<\pi$ is picked randomly from a Gaussian distribution centered around $\delta \theta=0$ with variance $\sigma^2=2 \Delta t/\tau_p(x)$ using the Box-Muller transform, $x$ being the instantaneous $x$-position. The next point is placed a distance $\Delta r_2=v_c(x) \Delta t$ in the $\theta_0+\delta \theta$ direction, this last step is repeated $N=t_{\rm tot}/\Delta t$ times to complete a trajectory representing a total time $t_{\rm tot}$. \ed{The time interval $\Delta t$ is chosen such that $\Delta t < \min_x(\tau_p(x))$; smaller than the smallest persistence time in the system. In all simulations shown here we chose $\Delta t=0.1$ hrs (corresponding to 6-minute intervals between measurements).} The substrate \edl{has a finite gradient region $x \in [-W,W]$}, with persistence time and velocity $\tau_{p,\rm{min}}$ and $v_{c,\rm{left}}$ at $x\leq -W$, and $\tau_{p,\rm{max}}$ and $v_{c,\rm{right}}$ for $x \geq W$. Both $\tau_p$ and $v_c$ transition linearly (but with variable steepness) \edl{controlled by $W$} between their \edl{max and min} values. Much like most experimental settings, the gradient region thus occupies only part of the system, and is flanked by uniformly \ed{rigid} regions to either side \ed{({\em i.e.}, the rigidity gradient changes discontinuously at the boundaries of the gradient region)}. We will always choose left to right to be the direction of increasing persistence but will, for demonstrational purposes, allow the velocity to decrease or increase from left to right. For each realization of the gradient region, on the order of $10^5$ trajectories are generated to obtain accurate averages.

\begin{figure}
\includegraphics[width=.8\columnwidth]{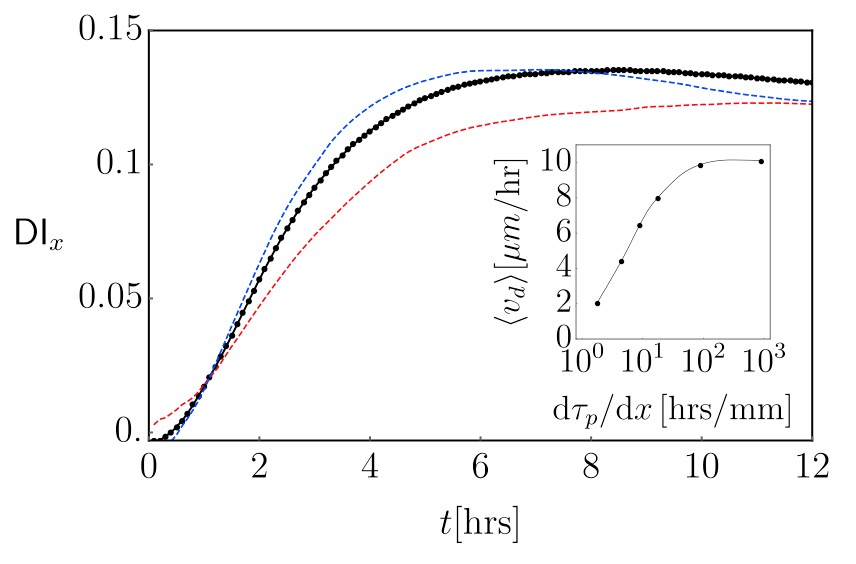}\caption{{\em Durotactic index as function of time}. Main figure: $x$-component of the durotactic index vs time for cells moving in a rigidity gradient, with $\tau_p$ increasing linearly from 0.2 to 2  \ed{hrs} over the $x$-range $[-0.1,0.1]$ mm. Averages computed over 5$\cdot 10^4$ trajectories \ed{(2D model)}. Black line, black dots: stiffness-independent velocity $v_c=50 \mu$m/hr everywhere. Red-dashed line: the same system, but with a velocity that {\em rises} with persistence; $v_c=20-80 \mu$m/hr accross the gradient region. Blue-dashed line: velocity {\em decreases} with persistence; $v_c=80-20 \mu$m/hr accross the gradient region. Inset: The \edl{average} velocity over the 12 hr window as a function of the gradient strength. All gradients had $\tau_p$ varying from 0.2 to 2 \ed{hrs}, but over different spatial ranges.} \label{Fig3}
\end{figure}

We assume, for now, that $v_c(x)\equiv v_c$; a constant (later on, we will briefly demonstrate that our findings are largely insensitive to increases or decreases in $v_c$ with stiffness). Our main finding is summarized in Fig. \ref{Fig1}: a gradient in persistence produces a soft-to-stiff flux of cells, and confers upon them, for typical values, an \edl{average} velocity up the stiffness gradient of 2-10 $\mu$m/hr. The origin of the effect is readily read off from Fig. \ref{Fig1} (a)-(c); PRW trajectories become asymmetric in the gradient region, and those trajectories that either depart up the rigidity gradient, or at some point in time first turn towards the stiff direction, travel further in the stiff direction, on average. \ed{As Fig. \ref{Fig1}(d) illustrates, this} leads to a nonzero $\langle x \rangle(t)$, and the \edl{average} velocity - over the $\sim 12$ hr course of a typical experiment, increases with increasing gradient steepness. \ed{We note, that in the limit of sufficiently small $\Delta t$, the dimensionless number \edl{${\sf V}=v_c\times ({\partial}\tau_p/\partial x)$} combines both parameters into a single quantity, and allows for a universal characterization of the durotactic motion. We choose to retain dimensional quantities to provide a sense of the magnitudes of velocities that may be expected in experimental settings. As Fig. \ref{Fig2} (a)-(d) shows, the asymmetry of a set of PRW trajectories on a substrate with gradient stiffness increases with time}. Fig. \ref{Fig2}(e) plots the probability distribution ${\sf P}(x,y)$ of finding a cell at position \ed{$(x,y)$} after $t=4$ hrs and shows the crucial statistical feature that gives rise to the nonzero center-of-mass motion. On the left, less persistent, side of the substrate the distribution resembles that of a diffusive process. On the right side, where motion is more persistent, a narrower peak moves outward at constant velocity. 

\begin{figure}[t]
\includegraphics[width=.9\columnwidth]{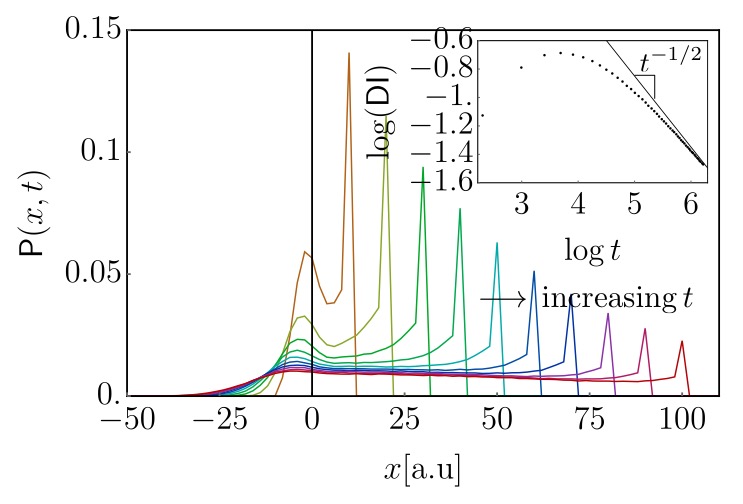}\caption{{\em Evolution of 1D inhomogeneous telegraph probability}. Probability distributions ${\sf P}(x,t)$ determined by direct integration of Eq. \ref{teleg} \ed{(1D model)}. The turning frequency $\lambda(x)$ decreased linearly from 0.4 to 0.02 over the $x$-interval $[-5,5]$. From left to right, we plot distributions for $t=10\dots 100$ with 10 unit time intervals. Clearly visible is the diffusive spreading on the left, vs. the wave-like propagation to the right. The inset shows the long-time $t^{-1/2}$ behavior of ${\sf DI}(t)$.} \label{Fig4}
\end{figure}

The net motion that results from differentially persistent PRW's executed in a stiffness gradient is reminiscent of the motion that chemotactic bacteria execute in, for instance, a gradient in nutrient concentration \cite{berg1972chemotaxis}. To be sure, in both cases an environmental gradient sets up a flux, but to what extent are these processes truly similar? Following \cite{mccutcheon1946vol, othmer1988models}, it is instructive to scrutinize the motility using a durotactic (vector-)index
\be
\vec \di(t)=\{\di_x(t),\di_y(t)\}\equiv \frac{\langle \vec r\rangle(t)}{v_c t}\, .
\ee 
\ed{In the case of variable cell-speed $v_c$, we compute $\vec \di(t)$ by dividing $\langle \vec r\rangle(t)$ by $r_{\rm path}=\int\!\! v_c(\vec r(t')) \rd t'$, the length of the path traveled up to time $t$.} For all - persistent and non-persistent - non-directional processes $\vec \di(t)=\vec 0$. For the gradients studied here $\di_y(t)=0$; we report only the $x$-component. In the main panel of Fig. \ref{Fig3}, we plot $\di_x(t)$ for a representative set of parameters (listed in the caption). The general behavior is, that $\di_x(t)$ initially rises, peaks at a few times the persistence time, and then slowly drops back down, proportional to $t^{-1/2}$ (cf., inset Fig. \ref{Fig4}). Fig \ref{Fig3} also shows, that this behavior remains qualitatively the same regardless of whether $v_c$ increases, decreases or stays the same through the gradient region. Since the DI is directly proportional to the \edl{average} velocity in the direction of the gradient, this is also the expected behavior for the \edl{average} velocity \ed{(see inset, Fig. \ref{Fig3})} which is thus a time dependent quantity for this \ed{process}. \edl{The large-time drop in $\di_x(t)$ is the result of walkers leaving the finite gradient region, and is generic; it also features when walkers exhibit a true directional bias in the same region, such as would occur for the 'run-and-tumble' behavior that underlies chemotaxis in, for instance, {\em E. coli} where it leads to a constant cellular drift velocity $v_d$ while in a chemical gradient region \cite{berg1972chemotaxis}. The short-time behavior, however, is completely different for these two processes: the biased run-and-tumble walk displays a $\di_x(t)$ rapidly saturating to its plateau value $v_d/v_c$, in contrast to a very gradual increase from zero for the differentially persistent walk. Thus, the presence or absence of a short-time regime of increasing $\di_x(t)$ is a reliable way to discriminate the motion we discuss here from a 'regular' taxis.} 

\vspace{1cm}

{\em 1D model and an inhomogeneous telegraph equation}. We map the process to one dimension by studying the dispersal of walkers on a line. The equivalent of a spatially dependent persistence, here, is a spatially dependent turning frequency $\lambda(x)$. Typical behavior in the presence of a gradient region is collected in Fig. \ref{Fig4} and confirms the dual behavior also seen in two dimensions: the softer side is diffusion-dominated while the more rigid side displays a wavelike propagation. To derive the appropriate continuum equation, we apply a similar approach to the one presented for uniform turning rates in \cite{othmer1988models}, and consider separately the two densities of left- and right movers; $\rho_-(x,t)$ and $\rho_+(x,t)$, normalized such that ${\sf P}(x,t)=\rho_+ + \rho_-$, is the total probability density. After a time step $\Delta t$, each walker reverses direction with a probability $\pi=\lambda(x) \Delta t$, or continues (with probability $1-\pi(x)$) along its prior direction. During each time step, it travels a distance $\Delta x=v_c \Delta t$. The densities $\rho_+$ and $\rho_-$ then obey

\begin{eqnarray}
\rho_+(x,t\!+\!\Delta t)&\!=\!&\left[1-\lambda(x\!-\!\Delta x)\Delta t\right]\rho_+(x\!-\!\Delta x,t)\nonumber\\
&+&[\lambda(x\!-\!\Delta x)\Delta t ]\rho_-(x\!-\!\Delta x,t)\, ,\\ 
\rho_-(x,t\!+\!\Delta t)&\!=\!&[\lambda(x\!+\!\Delta x)\Delta t ]\rho_+(x\!+\!\Delta x,t)\nonumber\\
&+&\left[1-\lambda(x\!+\!\Delta x)\Delta t\right]\rho_-(x\!+\!\Delta x,t)\, .
\end{eqnarray}
Expanding these two equations to first order in $\Delta x$ and $\Delta t$ and combining them using ${\sf P}=\rho_+ + \rho_-$ yields the following governing PDE
\be\label{teleg}
\partial_t^2 {\sf P}+2 \lambda(x) \partial_t {\sf P}=v_c^2\partial_x^2 {\sf P}\, .
\ee 
A spatially varying velocity may be included by replacing $v_c\to v_c(x)$. This inhomogeneous telegraph equation is also the appropriate \ed{model} to use for effectively one-dimension migration experiments. To connect with the two-dimensional case, we may identify $2\lambda(x)\simeq \tau_p^{-1}$. The two competing behaviors are readily recognized in the PDE; for large turning frequencies (i.e., short persistence times) the second order time derivative is dominated by the first order term, and diffusive behavior emerges. For low turning frequencies---highly persistent motion---a wave equation is recovered. This equation, supplemented with a specific form for the persistence gradient $\lambda(x)$, and the appropriate boundary conditions (generally, ${\sf P}(x,0)=\delta(x)$ and $\partial_t{\sf P}(x,0)=0$), allows one to compute averaged displacements as moments in this distribution.

{\em Conclusions and Outlook}. In this Letter, we demonstrate how a broadly reported feature of cellular motility---a dependence of the persistence of movement on the rigidity of the substrate---leads, without further assumptions, to universal soft-to-stiff motion on gradiented substrates. The motion is faster, on experimental timescales, for steeper gradients. For the type of motion we report here, the term durotaxis may be a bit of a misnomer. Following the suggestions laid out in \cite{tranquillo1990glossary}, the flux set up by gradients in the local, substrate-informed persistence is perhaps more accurately described as a (positional) kinesis---an "{\em almost instantaneous response induced by a purely positional signal}". That is, a non-directional change in behavior as opposed to the directional changes typical for chemotaxis. This distinction goes beyond semantics: it suggests that durotaxis in a stiffness gradient is not to be interpreted as the existence of a {\em preferred stiffness} for the cell, which it is purposefully migrating towards. Without dismissing the possibility that other mechanisms not considered here {\em could} lead to such properly durotactic motion, we show here that---at the very least to an extent that is worth determining in much greater detail---soft-to-stiff migration is an unavoidable consequence of stiffness-dependent persistence. \edl{The short-time behavior of $\di_x(t)$ may help distinguish this kinesis from properly tactic motion}. The generic nature of durokinesis suggests it as a potentially worthwhile mechanism to pursue in the development of instructive environments (for an early demonstration see, for instance, \cite{gray2003repositioning}); our results show that {\em any} stochastic, particulate system whose persistence is informed, locally, by some external parameter has the potential to harness this kinetic transport mechanism. 

\begin{acknowledgments}
This work was supported by funds from the Netherlands Organization for Scientific Research (NWO-FOM) within the program on Mechanosensing and Mechanotransduction by Cells (FOM-E1009M). We thank all FOM staff for 70 years of passionate commitment to Dutch physics, and Prof. James P. Butler for valuable discussions.
\end{acknowledgments}

\bibliography{references}

\end{document}